\documentstyle[preprint,aps]{revtex}
\tightenlines

\preprint{DPNU-99-14}
\begin{document}
\newcommand{\cms}{cm \( s^{-1} \)}
\newcommand{\cmt}{cm\(^{-3}\)}
\newcommand{\het}{\(^{3}\)He}
\newcommand{\hef}{\(^{4}\)He}
\newcommand{\alf}{Alfv\'{e}n}
\newcommand{\vct}[1]{\mbox{\boldmath\(#1\)}}
\newcommand{\parti}[2]{\frac{\partial #1}{\partial #2}}
\newcommand{\evect}{\vec{E}}
\newcommand{\bvect}{\vec{B}}
\newcommand{\kvect}{\vec{k}}
\newcommand{\xvect}{\vec{x}}
\newcommand{\vvect}{\vec{v}}
\newcommand{\pvect}{\vec{p}}
\newcommand{\gto}{\stackrel{>}{\sim}}
\newcommand{\lto}{\stackrel{<}{\sim}}
\newcommand{\valf}{v_{\rm A}}
\newcommand{\valft}{v_{\rm A}^{2}}
\newcommand{\drm}{{\rm d}}
\newcommand{\vsh}{v_{sh}}
\newcommand{\gsh}{\gamma_{sh}}
\newcommand{\vti}{v_{Ti}}
\newcommand{\vte}{v_{Te}}
\newcommand{\vfl}{v_{fl}}
\newcommand{\vflt}{v_{fl}^{2}}
\newcommand{\tj}{T_{j}}
\newcommand{\te}{T_{e}}
\newcommand{\ti}{T_{i}}
\newcommand{\mpr}{m_{p}}
\newcommand{\me}{m_{e}}
\newcommand{\mi}{m_{i}}
\newcommand{\mal}{m_{\alpha}}
\newcommand{\wpe}{\omega_{pe}}
\newcommand{\wpi}{\omega_{pi}}
\newcommand{\wce}{\omega_{ce}}
\newcommand{\wci}{\omega_{ci}}
\draft
\title{Electron acceleration to ultrarelativistic energies in a collisionless oblique shock wave}
\author{Naoki Bessho and Yukiharu Ohsawa}
\address{Department of Physics, Nagoya University, Nagoya 464-8602, Japan}
\date{\today}
\maketitle
\begin{abstract}
Electron motion in an oblique shock wave is studied by means of a one-dimensional, relativistic, electromagnetic, particle simulation code with full ion and electron dynamics. It is found that an oblique shock can produce electrons with ultra-relativistic energies; Lorentz factors with \( \gamma \gto 100 \) have been observed in our simulations. The physical mechanisms for the reflection and acceleration are discussed, and the maximum energy is estimated. If the electron reflection occurs near the end of a large-amplitude pulse, those particles will then be trapped in the pulse and gain a great deal of energy. The theory predicts that the electron energies can become especially high at certain propagation angles. This is verified by the simulations.
\end{abstract}

\section{Introduction}
\label{sec:level1}

\par
\indent
Electron acceleration has been an important issue in plasma physics and astrophysics. In solar physics, for instance, the acceleration of electrons as well as of ions has received a great deal of attention; in solar flares, gamma rays with energies of several tens of MeV, which are emitted by the bremsstrahlung from the high-energy electrons, are often observed \cite{rf:kan}, \cite{rf:yos}. In astrophysics, much more energetic electrons have been discussed. From the observations of X-rays and gamma rays, it is now believed that electrons with energies up to \( \sim \)100 TeV are produced by a shell-type supernova remnant SN 1006 \cite{rf:kp95}, \cite{rf:th98}. It is also believed that in the Crab Nebula high-energy electrons accelerated up to \( \sim \)100 TeV are produced by the pulsar wind \cite{rf:ts98}. Theories based on the Fermi acceleration model have been proposed to account for these high-energy phenomena \cite{rf:be}, \cite{rf:re}. As for laboratory plasmas, in an attempt to realize plasma-based accelerators, the electron acceleration in a relativistic space charge wave has been intensively studied by simulations and experiments \cite{rf:jcm}.
\par
\indent
Ion acceleration in a magnetosonic wave has been studied by many authors \cite{rf:bw}-\cite{rf:rt}. In a single-ion-species plasma, the large positive potential formed in the shock region can reflect some ions and give great energies to them \cite{rf:bw}-\cite{rf:lss}. In a multi-ion-species plasma such as space plasmas, some of the majority ions (hydrogen) can also gain energies by the same mechanism. Furthermore, the transverse electric field in the shock wave can accelerate all particles of all kinds of (minority) heavy ions to nearly the same speed \cite{rf:to95}, \cite{rf:to97}. In addition, in a turbulent plasma where particles can interact with many different large-amplitude magnetosonic pulses, some energetic ions can be further accelerated by a different mechanism \cite{rf:mbo}. These processes have been extensively studied by theory and particle simulation by many authors and have been applied to the production of high-energy particles in solar flares and interplanetary shocks \cite{rf:bw}-\cite{rf:to97}. 
\par
\indent
Strong electron acceleration in shock waves, however, has not been shown by particle simulations. In this paper we will theoretically and numerically study electron motion in a magnetosonic shock wave propagating obliquely to a magnetic field. We will show that an oblique shock can produce ultra-relativistic electrons. The preliminary result has been reported in Ref. \cite{rf:bmo}.
\par
\indent
In Sec.\ \ref{sec:level2}, we outline some basic properties of oblique shocks. In Sec.\ \ref{sec:level3}, we theoretically discuss the possibility of electron reflection. If electrons are reflected near the end of the main pulse, then after the reflection they can have great energies in the main pulse. 
\par
\indent
Several authors discussed stationary pulse solutions for the nonlinear magnetosonic waves \cite{rf:1}-\cite{rf:oh86}. However, if the magnetic field is rather strong so that \( \wce / \wpe \gto 1 \), where \( \wce \) is the electron cyclotron frequency and \( \wpe  \) is the plasma frequency, or if the amplitude is quite large, then the pulse propagation is not perfectly stationary \cite{rf:oh86b}, \cite{rf:ld2}; small-amplitude fluctuations are generated in and around the main pulse. Even though the potential is positive in the main pulse in a shock wave, the electron reflection studied in this paper can occur in such a non-stationary pulse. 
\par
\indent
In Sec.\ \ref{sec:level4}, we estimate the maximum energy of a reflected electron and discuss its dependence on plasma parameters. In Sec.\ \ref{sec:level5}, we further study the electron acceleration by using a one-dimensional (one space coordinate and three velocity components), relativistic, electromagnetic, particle simulation code with full ion and electron dynamics. It will be shown that some electrons can be reflected near the end of the main pulse and are trapped in the shock region. In the simulations, an appreciable number of electrons have energies \( \gamma > 100 \), where \( \gamma \) is the Lorentz factor. Further, we will examine the dependence of \( \gamma \) on the propagation angle and shock speed. It is verified that, as the theory predicts, there are special angles and shock speeds at which the electron acceleration is especially strong. Our work is summarized in Sec.\ \ref{sec:sum}. In Appendix A, we describe relations among physical quantities in a magnetosonic wave. In Appendix B, we give a rough estimate for the magnitude of parallel electric field \( E _{\parallel } \) on the basis of a simple physical picture.  

\section{Velocities and Fields in an Oblique Wave}
\label{sec:level2}
\par
\indent
We consider a magnetosonic wave propagating in the \(x\) direction with a speed \(v_{sh}\) in an external magnetic field in the (\(x, \ z\)) plane. We assume that the field quantities depend on \(x\) only; \(\partial / \partial y = \partial / \partial z = 0 \). Then, from the equation \(\nabla \cdot \vct{B} = 0 \), it follows that the  \(x\) component of the magnetic field is constant, \(B_x = B_{x0} \). The other components \(B_y \) and \(B_z \) are functions of \(x\), and in the pulse region  \(B_y \) as well as \(B_z \) can have finite values. In the wave frame where the time derivatives are zero (\(\partial / \partial t = 0\)), the electric field in the \(y\) direction is constant, \(E _y = E_{y0}\), and \(E _z\) is zero. Hence we have \(\vct{B} = (B_{x0}, \ B_y , \ B_z )\) and \(\vct{E} = (E_x , \ E_{y0}, \ 0 )\). In Sections \ref{sec:level2}-\ref{sec:level4}, theoretical analyses will be made mainly in the wave frame.
\par
\indent
The velocity of the guiding-center position of an electron, \(\vct{v}_{g} \), may be written as
\begin{equation}
\vct{v}_{g} = ( \vct{B} / B ) v_{\parallel} +  \vct{v}_{d}   ,
\label{eq:5}
\end{equation}
with \(\vct{v}_{d} \) being the drift velocity
\begin{equation}
\vct{v}_{d} =  c \vct{E} \times \vct{B} / B^2 - (c \mu / e ) ( \vct{B} \times \nabla B ) / B^2  .
\label{eq:10}
\end{equation}
Here \(c\) is the speed of light, \(- e\) is the electron charge (\(e > 0 \)), \(v_{\parallel} \) is the velocity parallel to the magnetic field, and \(\mu \) is the magnetic moment
\begin{equation}
\mu = \me v_{\perp} ^2 / ( 2 B ) ,
\label{eq:20}
\end{equation} 
with \( \me \) the electron mass and \( v_{\perp} \) the gyration speed perpendicular to the magnetic field. In the far upstream region, we have no \(\nabla B\)-drift. Also, the \(z\) component of the velocity averaged over all the electrons in a small volume element must be zero, \(\langle v_{{g}z0} \rangle = 0 \); the subscript 0 refers to the quantities in the far upstream region. Accordingly, from the \(z\) component of Eq. (\ref{eq:5}) we have the average parallel velocity
\begin{equation}
\langle v_{{\parallel}0} \rangle = \frac{c E_{y0}}{B_0 } \frac{B_{x0} }{B_{z0} } ,
\label{eq:30}
\end{equation}
in the wave frame (see Fig. \ref{fig:up}). Also, because \(\langle v_{{g}x0} \rangle = - v_{sh} \), the \(x\) component of Eq. (\ref{eq:5}) gives 
\begin{equation}
v_{sh} = - \frac{B_{x0} }{B_0 } \langle v_{{\parallel}0} \rangle - \frac{c E_{y0} B_{z0} }{B_{0} ^2 } .
\label{eq:40}
\end{equation}
Thus \(E_{y0} \) is related to the shock speed \(v_{sh}\) through 
\begin{equation}
E_{y0} = - v_{sh} B_{z0} / c .
\label{eq:50}
\end{equation}
For the definiteness, we assume that \( B _{x0} \), \( B _{z0} \) and \( v_{sh} \) are all positive; and thus \( E_{y0} < 0 \).
\par
\indent
The nonlinear wave theory tells us the wave structure. A shock wave will have a positive electric potential \( \varphi (x) \). The quantities \( \varphi \), \( B_z \), \(n_{i}\), and \(n_{e}\) have similar profiles \cite{rf:kak}-\cite{rf:oh86a} (see Appendix A); here \( n_{i} \), and \( n_{e} \) are the ion and electron densities, respectively. On the other hand, \( E_x \) and \( B_y \)  are proportional to the \( x \) derivatives of these quantities; for instance, \( E_x = - \partial \varphi / \partial x \). These relations among the quantities are obtained for small-amplitude waves. In the following theoretical analysis, we assume that the relations are also valid for large-amplitude waves. As we will see later in Fig. \ref{fig:eb},  simulation results also support this assumption.
\par
\indent
We denote by \( x_{m} \) the \( x \) position at which the potential takes its maximum value; hence, \( E_x (x_{m}) \simeq 0 \) and \( B_y (x_{m}) \simeq 0 \). For later use, we also note that the quantity \( B_z / B \) takes its maximum value at \( x = x_{m} \); that is, if \( B_z (x) \) is smaller than \( B_z ( x_{m} ) \), then 
\begin{equation}
\frac{ B_z ( x_{m} ) ^2 }{  B_{x0}  ^2  + B_z ( x_{m} ) ^2} > \frac{ B_z ( x ) ^2 }{  B_{x0}  ^2 + B_y ( x )  ^2 + B_z ( x ) ^2}  .
\label{eq:60}
\end{equation}
This can be proved by the following equation
\begin{equation}
\frac{ B _z ( x _m ) ^2 }{ B _{x0} ^2 + B _z ( x _m ) ^2 } -  \frac{ B _z ( x  ) ^2 }{ B _{x0} ^2 + B _y ( x ) ^2 + B _z ( x ) ^2 } = 
\frac{ [ B _z ( x _m ) ^2 - B _z ( x  ) ^2 ] B _{x0} ^2 + B _z ( x _m ) ^2 B _y ( x ) ^2 } { [ B _{x0} ^2 + B _z ( x _m ) ^2 ] [ B _{x0} ^2 + B _y ( x ) ^2 + B _z ( x ) ^2 ] } .
\label{eq:62}
\end{equation}
Because \(  B _z ( x _m ) \) is the maximum value of \( B _z  ( x ) \), the term \( [ B _z ( x _m ) ^2 - B _z  ( x ) ^2 ]  B _{x0} ^2 \) is positive. Hence the right-hand side of the above equation is positive.
\par
\indent
We know the stationary, finite-amplitude solutions for perpendicular waves \cite{rf:1}-\cite{rf:oh86}. The stationary solitary wave solution is valid when the magnetic field is weak, \( \wce / \wpe << 1 \), and when the wave amplitude is not large, \( 1 < M <2 \), where \( M \) is the \alf \ Mach number; \( M = \vsh / \valf \) with \( \valf \) the \alf \ speed, 
\( \valf ^2 = B _0 ^2 / ( 4 \pi n _i \mi ) \). 
Simulations and experiments show that large-amplitude pulses can propagate nearly steadily, even when the amplitudes are large or the magnetic fields are strong. Oblique pulses can also propagate nearly steadily. As the magnetic field becomes stronger or the wave amplitude is increased, non-stationary effects will become important \cite{rf:oh86b}, \cite{rf:ld2}.
\section{Electron Reflection}
\label{sec:level3}
\par
\indent
Because the potential is usually positive in the shock region, some of the ions can be reflected when they go up the potential. As we will see later, electrons can also be reflected; its mechanism is, however, different from that of ion reflection. We give a physical picture of the electron reflection. 
\par
\indent
From the nonrelativistic equation of motion for an electron
\begin{equation}
\me \frac{ \drm \vct{v} }{\drm t} =  - e \left( \vct{E} +  \frac{ \vct{v} \times \vct{B} }{c} \right)  ,
\label{eq:80} 
\end{equation}
we have an equation for the kinetic energy in the wave frame
\begin{equation}
\me (v^2 - v_0 ^2 ) / 2 =  e ( \varphi - \varphi _0 ) - e E_{y0} \int v_y \drm t  .
\label{eq:300} 
\end{equation}
Here we have used the relation \( E_z = 0 \). 
The velocity \( \vct{v}\) includes gyro, parallel, and drift motions. 
\par
\indent
We define length \( s \) as
\begin{equation}
{\rm d} s = (B / B _{x0} ) v_{gx} {\rm d} t  ,
\label{eq:90} 
\end{equation}
where \( v_{{g}x} \) is the \( x \) component of the electron velocity
\begin{equation}
v_{{g}x} = (B _{x0} / B ) v_{\parallel} + v_{dx}  ,
\label{eq:100} 
\end{equation}
(see Fig. \ref{fig:ds}). Thus \( \drm s /\drm t \) is given by
\begin{equation}
\frac{\drm s}{\drm t} = v_{\parallel} + \frac{B}{B _{x0}} v_{dx}  .
\label{eq:110} 
\end{equation}
We can interpret \( \drm s \) as an infinitesimal length along the field line corresponding to the length \( \drm x = v_{{g}x} {\rm d} t  \); \( \drm s = ( B / B _{x0} ) \drm x \) . In general the guiding-center velocity \( \vct{v} _{g} \) is not parallel to the magnetic field \( \vct{B} \). Hence the guiding center does not move along the field line. For a stationary, one-dimensional problem (\( \partial / \partial t = 0 \) and \( \partial / \partial y = \partial / \partial z = 0 \)), all quantities depend only on \( x \). As a result, for any stationary function \( f(x) \) we have 
\begin{equation}
\int _{x_1} ^{x_2} f(x) ( B / B _{x0} ) \drm x  = \int  _{ s _1} ^{ s _2} f(s) \drm s , 
\label{eq:111} 
\end{equation}
where \( s_1 \) and \( s_2 \) are the positions along the same field line; their \( x \) positions are \( x _1 \) and \( x _2 \), respectively. 
\par
\indent
If we define quantity \( F \) as
\begin{equation}
E_{\parallel} = - \drm F / \drm s , 
\label{eq:112} 
\end{equation}
then with the aid of (\ref{eq:90}) and the relation \( E_{\parallel} = \vct{E} \cdot \vct{B} / B \), we have
\begin{equation}
F = -\int \frac{ ( E_x B_{x0} + E_{y0} B _y  ) }{ B } \frac{ B }{B_{x0} } {\rm d} x  ,
\label{eq:120} 
\end{equation}
which can be written as
\begin{equation}
F = -\int \left( E_x - \frac{ \vsh  B _{z0} B _y  }{ c B_{x0} } \right) {\rm d} x  .
\label{eq:124} 
\end{equation}
Using the electric potential \( \varphi \) and the vector potential \( \vct{A} \) (\( \vct{B} = \nabla \times \vct{A} \)), we find
\begin{equation}
F = \varphi - ( \vsh / c ) (  B _{z0} / B _{x0} ) A_z   ,
\label{eq:130} 
\end{equation}
where \( A_z \) is the \( z \) component of the vector potential,
\begin{equation}
A_z = - \int B_y  \drm x  .
\label{eq:134} 
\end{equation}
The quantities \( \varphi \) and \( B_z \) have similar profiles, while \(B_y \) is proportional to \(\partial B_z / \partial x \) \cite{rf:kak}, \cite{rf:oh86a}. Hence  \( A_z \) and \( F \) have profiles similar to \( \varphi \).
\par
\indent
If we define quantity \( F _{l} \) in the laboratory frame (the subscript \( l \) denotes the laboratory frame) as
\begin{equation}
F _{l} = -\int  E _{l \parallel } ( B _l /  B _{lx0 } )  {\rm d} x _{ l} ,
\label{eq:126} 
\end{equation}
then for a stationary, one-dimensional problem, that is, for a case where wave profiles can be written as \( f ( x _l , t _l ) = f ( x _l - \vsh t _l )  \), we find  
\begin{equation}
F _{ l} = -\int \left( E _{l x }- \frac{ \vsh B _{ lz0 }  B _{ ly }  }{ c B_{lx0} } \right) {\rm d} x _{ l} .
\label{eq:126a} 
\end{equation}
One can readily show that it is related to \( F \) in the wave frame as 
\begin{equation}
F  = \gamma _{sh} F _{ l} , 
\label{eq:126b} 
\end{equation}
where \( \gamma _{sh} \) is defined as
\begin{equation}
\gamma _{sh} = ( 1 - \vsh ^2 / c ^2 ) ^{-1/2} . 
\label{eq:126c} 
\end{equation}
In Appendix B, we give a rough estimate of the maximum value of \( F \), Eq. (\ref{eq:a9}), found from qualitative physical considerations.
\par
\indent
Combining Eqs. (\ref{eq:300}) and (\ref{eq:130}), we eliminate \( \varphi \) 
\begin{equation}
\me v ^2 / 2 - \me v_{0}^2 /2 =  e (F - F_0 ) - ( e E_{y0}/ B_{x0} ) (A_z - A _{z0} + \int v_y B_{x0}  \drm t ) .
\label{eq:140} 
\end{equation}
By virtue of the definition of \( A_z \), (\ref{eq:134}), and the relation \( \drm x = v_x \drm t \), this equation can be written as
\begin{equation}
\me v^2 / 2 - \me v_{0}^2 /2 = e (F - F_0 ) + e( E_{y0} / B_{x0} ) \int ( v_{ x} B_y - v_{y } B_{x0} )   \drm t .
\label{eq:150} 
\end{equation}
Then, using the \( z \) component of the equation of motion, we find
\begin{equation}
\me v^2 / 2 - \me v_{0}^2 /2 = e (F - F_0 ) - \me c ( E_{y0} / B_{x0} ) (v_z - v_{z0}) .
\label{eq:160} 
\end{equation}
\par
\indent
In the drift approximation, the kinetic energy can be expressed as 
\begin{equation}
\me v^2 / 2 = \me (v_{\parallel} ^2 + v_{d} ^2 )/2 + \mu B . 
\label{eq:170} 
\end{equation}
Hence, averaging Eq. (\ref{eq:160}) over the electron cyclotron period, we have
\begin{eqnarray}
(  \me & / &  2 ) (v_{\parallel} ^2 + v_{d} ^2 ) + \mu B + \me c ( E_{y0} / B_{x0} ) v_{{g}z} - e F  
\nonumber \\
=  
( \me & / & 2 ) (v_{\parallel 0} ^2 + v_{\drm 0} ^2 ) + \mu B_0 + \me c ( E_{y0} /  B_{x0} ) v_{z0} - e F_0  . 
\label{eq:340} 
\end{eqnarray}
With the aid of the \( z \) component of Eq. (\ref{eq:5}), we can eliminate \( v_{{g} z}\) and obtain the following equation
\begin{eqnarray}
( \me / 2 ) ( v_{\parallel} & -  & v_{rv} ) ^2  +  K = e (F - F_0 ) - \mu ( B - B_0 )
\nonumber \\
+ \  ( \me / 2 )  (v_{\parallel 0} ^2 & + & v_{\drm 0} ^2 ) 
- \me c ( E_{y0} / B_{x0} ) ( v_{\drm z} - v_{z0} )  
- ( \me / 2 ) v _d ^2  ,
\label{eq:350} 
\end{eqnarray}
where \( v_{rv} \) and \( K \) are defined as
\begin{equation}
v_{rv} = - c E_{y0} B_z / ( B_{x0} B ) \ \ \ \ \ (>0),
\label{eq:350a} 
\end{equation}
\begin{equation}
K = - \me v_{rv}^2 /2 \ \ \ \ \ (<0) .
\label{eq:350b} 
\end{equation}
If we eliminate \( E_{y0} \) by substituting Eq. (\ref{eq:50}) in Eq. (\ref{eq:350a}), \( v_{rv} \) can be expressed as
\begin{equation}
v_{rv} = \vsh B_{z0} B_z / ( B_{x0} B ) .
\label{eq:350c} 
\end{equation}
We will see below that the velocity \( v _{gx} \) is reversed when \(v _{\parallel } = v _{rv} \). In the pulse region, the quantity \( e (F - F_0 ) \) can be much larger in magnitude than the other terms on the right-hand side of Eq. (\ref{eq:350}) (see Appendix B). We denote the left-hand side of Eq. (\ref{eq:350}) by \( E _e \), i.e., \( E _e  = ( \me / 2 ) ( v_{\parallel} - v_{rv} ) ^2 + K \). As shown in Fig. \ref{fig:ekin}, at a fixed point \( x \), \( E _e \) has its minimum value \( K \) when \( v_{\parallel}= v_{rv} \). It is negative in the region
\begin{equation}
0 < v_{\parallel} < 2 v_{rv} . 
\label{eq:360} 
\end{equation}
\par
\indent
Because \( v _{dx} =  c E _{y0} B _z / B ^2 \), it follows from Eq. (\ref{eq:100}) that \( v _{gx} = 0 \) if \( v _{\parallel } = v _{rv} \). As can be seen from Eq. (\ref{eq:30}), most of the electrons have negative parallel velocities in the far upstream region. Accordingly, we suppose that the initial parallel velocity is negative, \( v_{\parallel} < 0 \). If \( v_{\parallel} \) gradually increases and exceeds \( v_{rv} \) at a certain position, then the \(x\) component of the guiding center velocity, \( v_{{g}x} \), changes from negative to positive values there, i.e., it is reversed.
\par
\indent
As an electron moves, i.e., as \( x \) changes, the quantities \( K \), \( F \), \( B \), etc. also vary. At any point \( x \), \( K \) is the minimum value of the left-hand side of Eq. (\ref{eq:350}). If there is a region where the values of the right-hand side of Eq. (\ref{eq:350}) become smaller than \( K \), then the electron cannot enter there; it will be reflected. We recall that \( \varphi \), \( F \), and \( B_z \) have similar profiles and that at \( x = x_{m} \) the quantity  \((B_z / B )\) has a maximum value [see Eq. (\ref{eq:60})]. Because \( K \) is proportional to \( (B_z / B ) ^2 \), it will be quite small (\( |K| \) is large) around the point \( x = x_{m} \), and \( F \) is large. In front of or behind the main pulse, \( F \) is relatively small and \( K \) is relatively large; \( e (F - F _0 ) \) could be smaller than \( K \). Thus, if the electron reflection occurs, it will be in such regions.
\par
\indent
In quasi-perpendicular shocks, where \( ( B_{z0} / B_{x0} ) >> 1 \), \( v_{rv} \) is large. Hence \( K \) has large negative values. Thus the electron reflection in this mechanism will be difficult.
\par
\indent
Figure \ref{fig:orb} shows a schematic diagram of the trajectory of guiding center. Here it is assumed that the reflection takes place at point D; the dotted line shows the orbit of a passing electron. As an electron moves from point A to C, it moves to the negative \( y \) direction because of the \( E_x \times B _z \) drift. It gains kinetic energy \( \Delta E _1 \) from the electric potential
\begin{equation}
\Delta E_1 = e \varphi (x_{C}) - e \varphi (x_{A}) \ \ \ \ (>0) .
\label{eq:365a} 
\end{equation}
At the same time, it loses energy \( \Delta E _2 \) because of the electric field \( E_{y0}\)
\begin{equation}
\Delta E_2 =  - e  E_{y0} (y_{C} - y_{A})  \ \ \ \ (< 0)  .
\label{eq:365b} 
\end{equation}
The net change in the energy is therefore
\begin{equation}
\Delta E = \Delta E_1 + \Delta E_2 .
\label{eq:365c} 
\end{equation}
Even though the magnitudes of \( \Delta E_1 \) and \( \Delta E_2 \) are quite large, they almost cancel when an electron moves with drift approximation; in particular, in a perpendicular pulse they have exactly the same magnitude and \( \Delta E = 0 \) \cite{rf:oh89}. However, if an electron is reflected and moves from D to E, then it would gain energies from both \( E_x \) and \( E_{y0} \). As a result, the increase in energy is
\begin{equation}
\Delta E = \Delta E _1 + \Delta E _3  ,
\label{eq:365d} 
\end{equation}
where \( \Delta E _3 \) is defined as
\begin{equation}
\Delta E_3 =  - e  E_{y0} (y_{E} - y_{A})  \ \ \ \ (>0)  .
\label{eq:365e} 
\end{equation}

\section{Estimate of Energy Increase}
\label{sec:level4}
\par
\indent
Next, we will obtain the maximum energy of a reflected electron. Because it can have quite high energy, we will use the relativistic equation of motion
\begin{equation}
\me \frac{ \drm ( \gamma \vct{v} ) }{\drm t} =  - e \left( \vct{E} +  \frac{ \vct{v} \times \vct{B} }{c} \right)  ,
\label{eq:178} 
\end{equation}
where \( \gamma \) is the Lorentz factor. For the stationary, one-dimensional system being considered, this can be integrated to give the energy conservation equation
\begin{equation}
\me \gamma c ^2  - \me \gamma _0 c^2  =  e ( \varphi - \varphi _0 ) -e E_{y0} \int v_y \drm t .
\label{eq:180} 
\end{equation}
This is a relativistic form of Eq. (\ref{eq:300}). Again, using the quantities \( F \), \( A_z \), and the equation of motion, we can rewrite Eq. (\ref{eq:180}) as
\begin{equation}
\me \gamma c ^2  - \me \gamma _0 c^2  =  e (F - F_0 )  - \me c ( E_{y0}/ B_{x0} ) ( \gamma v_z - \gamma _0 v_{z0} ) .
\label{eq:190} 
\end{equation}
\par
\indent
If we define quantity \( h \) as
\begin{equation}
h = \me c ^2 - \me v_{sh} B_{z0} v_{z}  / B_{x0}       ,
\label{eq:192} 
\end{equation}
then from Eq. (\ref{eq:190}) we have
\begin{equation}
\gamma  = [ e ( F - F_0 ) + h_0 \gamma _0 ] / h       .
\label{eq:200} 
\end{equation}
Equation (\ref{eq:192}) indicates that \( h \) is positive if \( B_{z0} / B_{x0}  \) is of order unity. If \( B_{z0} / B_{x0}  \)  is much greater than unity, then \( h \) can be negative. In addition, Eq. (\ref{eq:200}) shows that, if the initial value of \( h \) is positive, \( h_0 > 0 \), then \( h \) is always positive in the region where (\( F- F_0 \)) is positive.
\par
\indent
We show in Fig. \ref{fig:fh} schematic diagram of the function [\( e ( F - F_0 ) + h_0 \gamma _0 \)]. We suppose that \( F \) has a maximum value at \( x = x _{m}\) and has a minimum value right behind the main pulse. If the reflection occurs, it would be in this dip, as suggested in the previous section. The reflection point will be denoted by \( x _{r} \). In the top panel this function \( e ( F - F_0 ) + h_0 \gamma _0 \) is always positive. In the second one it becomes negative right behind the main pulse; however its maximum value is positive  (\( h _0 \) can be either positive or negative). In the bottom one it is always negative. For the present acceleration mechanism, the top panel is the most important; the strong acceleration takes place in this case.
\par
\indent
Such dips would not be present in the perfectly stationary solutions \cite{rf:1}-\cite{rf:oh86a}. In large-amplitude magnetosonic waves, however, non-stationary dips can be generated \cite{rf:oh86b}, \cite{rf:ld2}. We here just assume that there is a dip behind the main pulse and discuss its effects on particle orbits. 
\par
\indent
First we consider the top panel in Fig. \ref{fig:fh}: An electron with positive \( h_0 \) is reflected at a certain point near the end of the main pulse. We will show that, after the reflection, \( h \) has its minimum (positive) value at  \( x = x_{m} \); hence \( \gamma \) has its maximum value there. 
\par
\indent
As mentioned earlier, the parallel velocity \( v_{\parallel} \) is already positive when an electron is reflected, i.e., when \( v_{{g} x} \) is reversed, even if its initial value is negative. After the reflection, \( v_{\parallel} \) will be increased by the (positive) parallel electric force. Because \( E _{\parallel } \) changes its sign at \( x = x_{m} \), \( v_{\parallel} \) will have a maximum value at \( x = x_{m} \) (unless \( v_x \) changes to negative values again before reaching the point \( x_{m} \) ). The parallel velocity will become the dominant component in the electron velocity, \( \vct{v} \sim v_{\parallel} \vct{B} / B \). The quantity \( B_z / B \) as well as \( \varphi \) and \( B_z \) has a maximum value at this point.  Thus \( v_z \), which can be approximated as \( v_z \sim v_{\parallel} B_z / B \), will also have the maximum value there. Because \( h_0 \)  and (\( F_{m} - F_0 \)) are both positive, \( h \) is positive at \( x = x_{m} \). In addition, Eq. (\ref{eq:192}) indicates that \( h \) decreases (draws closer to zero) as \( v_z \) increases. Hence, \( h \) of a reflected electron will have a minimum positive value at \( x = x_{m} \). Its energy \( \gamma \) will therefore have a maximum value there. If \( h \) can be close to zero, then \( \gamma \) would become quite large.
\par
\indent
If \( B_{x0} \) is small and \( v_{z0} \) has a large positive value, then \( h_0 \) becomes small (can be even negative). We now consider such particles; the second panel in Fig. \ref{fig:fh}. Also in this case, \( v_z \) will have larger values after the reflection; thus \( h \) will decrease. At the reflection point in the dip, the quantity \([ e ( F  -  F_0 ) + h_0 \gamma _0 ] \) is negative. Consequently, \( h ( x _{r} ) < 0 \). If the reflection occurs, the electron would move in the positive \( x \) direction. Because \( v _z \) increases, \( h \) will further decrease. Hence, the particle would never be able to penetrate the region again where \( h \gamma > 0 \). The strong acceleration in this mechanism, therefore, will not take place.
\par
\indent
Next, we consider a case where the quantity [\( e(F - F_0 ) + h_0 \gamma _0 \)] is always negative (the bottom panel in Fig. \ref{fig:fh}). Then, \( h \) is always negative. If the reflection occurs at \( x = x _{r} \), then \( |h| \) would increase while \( |e(F - F_0 ) + h_0 \gamma _0 | \) would decrease. Hence, even if the particle reaches the point \( x _{m} \), \( \gamma \) would not have an extremely large value. Therefore, the strong acceleration is not expected. 
\par
\indent
We can estimate the maximum value \( \gamma _{m} \) in the wave frame by substituting the maximum value of \( F \), Eq. (\ref{eq:a12}), in Eq. (\ref{eq:200}). The maximum \( \gamma \) in the laboratory frame can be obtained by the Lorentz transformation as
\begin{equation}
\gamma _{lm} =  \gamma _{sh} (1 + v_x  v_{sh} / c ^2 ) \gamma_{m} \simeq    \gamma _{sh} \gamma_{m} .
\label{eq:210} 
\end{equation}
\par
\indent
Now let us consider a case where \( h \) becomes quite small, \( h \simeq 0 \), at \( x = x_{m} \). There the electron speed would be close to the speed of light \( c \).  Hence, roughly, we have \( v_z \sim c B_z / B \). The quantity \( h \) can then be expressed as 
\begin{equation}
h =  \me c ^2  \left( 1 - \gamma _{sh} \frac{ v_{sh} }{ c } \frac{ B_z }{B} \tan \theta \right)      ,
\label{eq:220} 
\end{equation}
where \( \tan \theta = B_{{l}z0}/ B_{{l}x0} \). Here we have used the relations \( B_{z0} = \gamma _{sh} B_{{l}z0} \) and \( B_{x0} = B_{{l}x0} \). Because \( B_z \) becomes quite large in the shock region, we may put as \( B_z / B \sim 1 \). Equation (\ref{eq:220}) then suggests that the electron energy can become extremely large when
\begin{equation}
\gamma _{sh} ( v_{sh} / c ) \tan \theta = 1      ,
\label{eq:230} 
\end{equation}
or, equivalently
\begin{equation}
c \ \cos \theta = v_{sh}    .
\label{eq:230a} 
\end{equation}

\section{Simulation Studies}
\label{sec:level5}
\subsection{Simulation Method}
\label{sec:sub1}
\par
\indent
We further study the electron acceleration by using a one-dimensional (one space coordinate and three velocity components), relativistic, electromagnetic, particle simulation code with full ion and electron dynamics \cite{rf:lldc}. The total system length is \( L_x = 4096 \Delta _{g} \) for most of the simulation runs, where \( \Delta _{g} \) is the grid spacing. All lengths and velocities in the simulations were normalized to \( \Delta _{g} \) and \( \wpe \Delta _{g} \), respectively, where \( \wpe \) is the spatially averaged plasma frequency.
\par
\indent
We use a bounded plasma model. The plasma is limited to the region \( 400 < x < 3696 \). The particles are specularly reflected at \( x = 400 \) and at \( x = 3696 \). The radiation leaving the plasma region is absorbed in the vacuum regions, \( 0 < x < 400 \) and \( 3696 < x < 4096 \). Thus the electromagnetic interactions between the two plasma boundaries through the vacuum regions are made negligibly small \cite{rf:oh85}.
\par
\indent
In most of the plasma region, i.e., in \( 500 < x < 3696 \), each particle species initially has a uniform density and has a Maxwellian velocity distribution function. In the small region, \( 400< x < 480 \), the plasma density is four times as high as that in the main region. That is, the initial plasma density can be written as
$$
n _{e}  = 4 n _0 \ \ \ \ \ \mbox{for} \ \  400 < x < x_1  ,
$$
$$
n _{e}   = 4 n _0 \exp [ - ( x - x_1 ) ^2 / ( 2 d ^2 ) ] \ \ \ \ \ \mbox{for} \ \  x_1 < x < 500 ,
$$
$$
n _{e} = n _0 \ \ \ \ \ \mbox{for} \ \  500 < x < 3696.
$$
Here  \( x_1 = 480 \)  and \( d = 12 \). In the high-density region, \(400 < x < 480 \), particles have shifted Maxwellian velocity distribution functions with average velocity \( \vct{v} _{ps } \); \( f _j \sim \exp [ - (\vct{v} - \vct{v} _{ps } ) ^2 / ( 2 v _{{T} j } ^2 ) ] \), where the subscript \( j \) refers to particle species. These particles act as a piston. They push the neighboring particles and excite a shock wave. We can change the shock strength by changing the magnitude of \( \vct{v} _{ps } \); the velocity \( \vct{v} _{ps } \) is perpendicular to the external magnetic field so that no particles have extremely large parallel speeds initially. More detailed description about the simulation code can be found in Ref. \cite{rf:oh85}.
\par
\indent
The simulation parameters are as follows. The number of simulation particles is \( N_{i} = N_{e} = 262,144 \). The ion-to-electron mass ratio is \( \mi / \me = 100 \). The ratio \( \wce / \wpe \) is \( \wce / \wpe = 3.0 \) in the far upstream region. The electron and ion thermal velocities are \( \vte / c = 0.38 \) and \( \vti / c = 0.01 \), respectively. The \alf \ speed is \( \valf / c = 0.3 \). The electron skin depth is \( c / \wpe = 4 \Delta _g \). The time step is sufficiently small, \( \wpe \Delta t = 0.02 \), so that \( \Delta t  \) is much smaller than the plasma and cyclotron periods even in the shock region. As in the theoretical model, the external magnetic field is in the (\( x , \ z \)) plane, and waves propagate in the \( x \) direction.

\subsection{Simulation Results}
\label{sec:sub2}
\par
\indent
First, we study a shock with the propagation angle \( \theta = 45 ^{\circ } \); i.e., \( B _{x0} = B _{z0} \). Figure \ref{fig:mag} shows profiles of \( B_z \) in the shock wave at various times. Here, the magnetic field is normalized to \( B _{z0} \). The Mach number is observed to be \( M= 2.3 \), i.e., the shock speed is \( \vsh = 2.3 \valf \). The maximum value of \( B _z \) is  4\( \sim \)5 times as large as \( B _{z0} \). The main pulse propagates nearly steadily. However, the profile is not perfectly stationary. 
\par
\indent
Figure \ref{fig:eb} displays field profiles at \( \wpe t = 680 \); here the magnetic and electric fields are both normalized to \( B _0 \). As the theory predicts \cite{rf:kak}, \cite{rf:oh86a}, \( E _y \) and \( B _z \) have similar profiles, while \( E _x \), \( E _z \), and \( B _y \) are quite small in magnitude at the point where \( B _ z \) has its maximum value. Figure \ref{fig:phase} shows phase space plots of electrons at \( \wpe t = 680 \). In the shock region, ultra-relativistic electrons are produced. The maximum value of Lorentz factors is \( \gamma \simeq 130 \). We also see that \( p _x \) and \( p _z \) have greater values than \( p _y \), which is consistent with the picture that the parallel velocity is dominant in the motion of high-energy electrons. From Figs. \ref{fig:eb} and \ref{fig:phase}, we see that some electrons are reflected near the end of the main pulse and that the high-energy electrons are present in the shock region.  
\par
\indent
Figure \ref{fig:efp} shows profiles of \( E _{\parallel }\),  \( F \), and the electric potential  \( \varphi \) at \( \wpe t = 650 \) and  at \( \wpe t = 680 \). The quantities \( \tilde{ F } \) and \( \tilde{ \varphi } \) denote \( e F / ( \me c ^2 ) \) and \( e \varphi / ( \me c ^2 ) \), respectively. The quantities \( F \) and  \( \varphi \) were obtained in the simulation by the following equations
\begin{equation}
F (x) = - \int ^{x} E _{\parallel} ( B / B _{x0} ) \drm x ,  
\label{eq:400} 
\end{equation}
\begin{equation}
\varphi (x) = - \int ^{x} E _x \drm x .  
\label{eq:410} 
\end{equation}
As mentioned earlier, roughly speaking, the parallel electric field \( E _{\parallel } \) is positive in the region where \( \partial B _z / \partial x \) and \( \partial F / \partial x \) are negative. It is negative in the region where they are positive. Comparing the plots at different times, we see that the wave profiles are not perfectly stationary. They vary with time with small-amplitudes. In particular, as shown in the plot at \( \wpe t = 680 \), \( F \) sometimes becomes negative near the end of the main pulse, where the electron reflection can take place.
\par
\indent
Next, we will study trajectories of electrons; we will find that it is the reflected electrons that gain great energies. Figure \ref{fig:xyzg} displays time variations of (\( x - v_{sh} t \)), \(y\), \(z\), and \(\gamma \) of electrons in a shock with \(v_{sh} = 1.75 \valf \); (\( x - v_{sh} t \)) is the \( x \) position in the wave frame. The lengths are normalized to the electron skin depth \( c / \wpe \). (Here the system size is a half of that in the previous run; the other parameters such as \(N _{e} / L _x \) are the same.) Two typical examples are presented here. The thick lines show an electron that was accelerated, and the thin lines represent an electron that was not accelerated; even the low-energy electron here has energy \(\gamma \simeq 10 \) in the shock region and \(\gamma \simeq 5\) behind the shock. In all of the four panels, the low- and high-energy electrons both have small-amplitude, cyclotron oscillations with short periods, \( \wpe t \lto  20\); note that the cyclotron period depends on \(\gamma \) as well as the field strength \(B\). At the same time, the accelerated electron has a large-amplitude, long-period oscillation with period \( \wpe  t \simeq 170\); it begins when the particle enters the shock, \( \wpe t = 180\), and continues until the end of this simulation run. (Simulations show that this period increases with  \( \gamma \). To observe this oscillation as many times as possible, we have chosen here the case where \( \gamma \) is not so large.) For comparison with Fig. \ref{fig:orb}, we show the times corresponding to points B, C, D, and E.  From the long-period oscillation of (\(x - v_{sh} t \)) in the top panel, we see that the electron that will become high energy is reflected and then trapped by the shock wave. On the other hand, the quantity (\(x - v_{sh} t \)) of the low-energy electron keeps decreasing, which means that this particle passes through the shock region without strong interactions. The second panel shows that the \(y\) position of the accelerated electron also oscillates, while the third one shows that \(z\) increases rapidly after the reflection. The average value of velocity \(v_z \) over the period from \(\wpe t = 220 \) to 400 is \(v_z / c \simeq 0.6\), and the value of \(v_z \) at maximum \(\gamma \) is \(v_z / c \simeq 0.8\). The maximum \(\gamma \) in the bottom panel is \(\gamma \simeq 40\). 
\par
\indent
By the Lorentz transformation of these data, we can find \( \vct{v} \) and \( \vct{B} \) in the wave frame. At the time and position of maximum \( \gamma \), they are \( \vct{v} =\) \(c ( 0.27,\ 0.13, \ 0.95)\) and \( \vct{B} =\) \(B_0 ( 0.70,\ 0.16, \ 2.56)\). This confirms our statement in the previous section that \( v_{\parallel } \) is the dominant component in \( \vct{v} \) there; i.e., \( v_z \simeq v_{\parallel } B_z / B \).
\par
\indent
According to the theory, (\ref{eq:230a}), electron energies can be extremely high at some angles and shock speeds. To examine this prediction, we have carried out several simulations with different values of the propagation angle \( \theta \), keeping other parameters unchanged; \( L _x = 4096 \Delta _{g} \), \( \wce / \wpe = 3.0 \), and \( \vsh \simeq 2.1 \valf \). Figure \ref{fig:angle} shows the maximum value of \( \gamma \) as a function of the propagation angle \( \theta \). Here the electron energy takes the highest value at the angle \( \theta \simeq 52 ^{\circ } \).  For these simulation parameters, Eq. (\ref{eq:230a}) predicts that \( \gamma \) takes a peak value at \( \theta \simeq 53 ^{\circ } \). Thus the theory and simulation are in good agreement. Further, we carried out simulations with different values of the shock speed \( \vsh \); the propagation angle is fixed to be \( \theta = 45^{\circ }\). Figure \ref{fig:vsh} shows the maximum value of \( \gamma \) as a function of the shock speed \( v_{sh} \). The observed values have a peak at \( \vsh / c \simeq 0.7 \). For these parameters, Eq. (\ref{eq:230a}) predicts that \( \gamma \) takes large values for \( \vsh /c \simeq 0.71 \). Again, the theory and simulation are in good agreement. 

\section{Summary}
\label{sec:sum}
\par
\indent
\ 
\par
\indent
We have studied electron motion in a shock wave propagating obliquely to a magnetic field. First, we analytically discussed electron motion in an oblique shock wave. It is pointed out that, if an electron is reflected near the end of a large-amplitude pulse, it would gain a great deal of energy from the electric field formed in the wave. The condition for the reflection is then examined. Further, the maximum energy of a reflected electron is estimated, and its dependence on plasma parameters is discussed. Next, we investigated the shock evolution and associated electron acceleration by using a one-dimensional, relativistic, electromagnetic, particle simulation code with full ion and electron dynamics. It was shown that an oblique shock can produce ultra-relativistic electrons; Lorentz factors with \( \gamma \gto 100 \) have been observed. As the theory predicts, at certain propagation angles electron energies become extremely high. The electron reflection and resultant acceleration takes place when the wave amplitude is large and the magnetic field is rather strong, \( \wce \gto \wpe \). 
\par
\indent
In the present paper, strong electron acceleration has been demonstrated. As future work, it would be desirable to develop further the quantitative theory for large-amplitude oblique waves. The evaluation of the electric field strength along the magnetic field will be especially important, which will enable us to estimate the maximum energy more accurately. From the view point of nonlinear wave propagation, motion of trapped electrons and their effects on the wave evolution are also quite interesting.
\appendix
\section{Relations among Quantities}
\par
\indent
We here describe relations among the physical variables, using the approximation appropriate for low-frequency magnetosonic waves with small amplitudes (for the details of the calculations, see Ref. \cite{rf:oh86a}). We assume quasi neutrality, \( n \simeq n _i  \simeq n _e \) and use the following stretched variables 
\begin{equation}
\tau = \epsilon ^{3/2} t ,
\label{eq:rq1}
\end{equation}
\begin{equation}
\xi = \epsilon ^{1/2} ( x - v _{p0} t ) ,
\label{eq:rq2}
\end{equation}
(\( v _{p0} \) is the wave propagation speed in the long-wavelength limit in a finite beta plasma) and expansion
\begin{equation}
n  = n _0 + \epsilon n _1 + \epsilon ^{2} n _2 + ... ,
\label{eq:rq3}
\end{equation}
\begin{equation}
E _y = \epsilon E _{y1} + \epsilon ^{2} E _{y2} + ... ,
\label{eq:rq4}
\end{equation}
\begin{equation}
B _z = B _{z0} + \epsilon B _{z1} + \epsilon ^{2} B _{z2} + ... ,
\label{eq:rq5}
\end{equation}
then, we can express the lowest order perturbations in terms of \( n_1 \) as
\begin{equation}
B _{z1} / B _0  = c E _{y1} / ( B _0 v _{p0} ) = [ ( v _{p0} ^2 - c _s ^2 ) / \valf ^2 \sin \theta ] ( n _1 / n _0 )  ,
\label{eq:rq6}
\end{equation}
\begin{equation}
\frac{ c E _{z1} }{ B _0 } = - \frac{ v _{p0} B _{y1}}{ B _0 } = \frac{ ( \wce  - \wci  ) }{ \wce \wci } \frac{ ( v _{p0} ^2 - c _s ^2 ) v _{p0} ^2 }{ ( v _{p0} ^2 - \valf  ^2  \cos ^2 \theta ) } \frac{ \cot \theta }{ n _0 } \frac{ \partial n _1 }{ \partial \xi } .
\label{eq:rq7}
\end{equation}
Here \( c _s \) is the sound speed, \( c _s ^2 = ( \Gamma _e p _{e0} + \Gamma _i p _{i0}) / (n _0 \mi ) \) with \( \Gamma _j \) (\(j = e \) or \( i \)) the specific heat ratio and \(p _{j0} \) the equilibrium pressure.  
We can obtain KdV equation by proceeding to the next order, \( O ( \epsilon ^{5/2} ) \).
\par
\indent
The relations (\ref{eq:rq6}) and (\ref{eq:rq7}) are valid, even when the wave profile is not soliton-like. That is, perturbations propagating in the same direction with \( v _x \sim v _{p0} \) have the relations (\ref{eq:rq6}) and (\ref{eq:rq7}).
\section{Magnitudes of \( E _{\parallel }\) and \( F \)}
\par
\indent
For a nonrelativistic, small-amplitude oblique magnetosonic wave, the electric field parallel to the magnetic field, \( E _{\parallel } \), has been analytically obtained \cite{rf:oh88}. However, for a large-amplitude wave, it will be quite difficult to obtain \( E _{\parallel } \) in a rigorous manner. Here, on the basis of a simple physical picture, we will give a rough estimate of the parallel electric field \( E _{\parallel } \) in a relativistic, large-amplitude, oblique shock wave. Further, by using it, we will obtain the maximum value of the quantity \( F \). 
\par
\indent
We consider in the wave frame. In the fluid model, the equation of motion may be written as 
\begin{equation}
m _j \frac{ \drm ( \gamma _j \vct{v} _j ) }{ \drm  t }  =  q _j \vct{E} + q _j \frac{ \vct{v} _j  \times \vct{B} }{c} .
\label{eq:a1}
\end{equation}
Here the pressure term is neglected. The subscript \(j\) refers to ions \(j =  i \) or electrons \(j =  e\). The time derivative \( \drm / \drm t \) is defined as
\begin{equation}
\frac{ \drm  }{ \drm  t }  = \frac{ \partial }{ \partial t } + \vct{v} _j \cdot \nabla \ \ \  .
\label{eq:a2}
\end{equation}
We denote by \( \vct{b} \) the unit vector along the magnetic field. Then multiplying Eq. (\ref{eq:a1}) by \( \vct{b} \), we have
\begin{equation}
m _j \frac{ \drm ( \gamma _j \vct{b} \cdot \vct{v} _j ) }{ \drm  t } - m _j \gamma _j \vct{v} _j \cdot \frac{ \drm \vct{b} }{ \drm t} = q _j E _{\parallel } .
\label{eq:a3}
\end{equation}
\par
\indent
Let \( t _1 \) be the time when a fluid element is at the leading edge of the shock (the location where the shock profile begins to sharply rise) and \( t _2 \) be the time when it is at the point \( x _{m} \); \( x _{m} \) is the location where the electric potential and magnetic field take their maximum values. Then, integrating from \( t _1 \) to \( t _2 \), we find for the ions 
\begin{equation}
\mi  (  \gamma _{i2} \vct{b} _2 \cdot \vct{v} _{i2}  - \gamma _{i1} \vct{b} _1 \cdot \vct{v} _{i1} ) - \mi \langle  \gamma _i \vct{v} _i \rangle  \cdot ( \vct{b} _2 - \vct{b} _1 ) \simeq e E _{\parallel } \frac{ \Delta }{ | \langle v _{ix} \rangle | } .
\label{eq:a4}
\end{equation}
Here the bracket indicates the mean value over the time period (\( t _2 - t _1 \)), and \( \Delta \) is the shock width. We have a relation (\( t _2 - t _1 ) \simeq   \Delta / | \langle v _{ix} \rangle  | \). For oblique shocks, \( \Delta \) will be of the order of the ion inertial length, \( \Delta \simeq c / \wpi \). The velocity at time \( t _1 \) is \( \vct{v} _{i1} = - \vsh \vct{e} _x \), where \( \vct{e} _x \) is the unit vector in the \( x \) direction. The velocity \( v _{ix} \) will be slowed down by the longitudinal electric field. Thus \( v _{ix2} \) will be smaller in magnitude than \( v _{ix1} \). Because \( B _z \) is large at the point \( x _{m} \) and \( B _y ( x _{m} ) \sim 0 \), the unit vector \( \vct{b} _2 \) is nearly parallel to the \( z \) direction. Consequently, the inner product (\( \vct{b} _2 \cdot \vct{v} _{i2} \)) will be quite small; here we neglect the \( z \) component of \( \vct{v} _{i2} \). Since the \( x \) component of \( \vct{b} _1 \) is given by \( B _{x0} / B _0 \), we have  
\begin{equation}
\mi ( \gamma _{i1} \vsh + \langle  \gamma _{i} v _{ix} \rangle  ) B _{x0} / B _0  \simeq  e E _{\parallel } \Delta / | \langle v _{ix} \rangle |  .
\label{eq:a5}
\end{equation}
This gives the strength of the parallel electric field as
\begin{equation}
e E _{\parallel } \simeq \frac{ \mi  \vsh ^2 }{ \Delta } \frac{ | \langle v _{ix} \rangle | }{ \vsh } \left( \gamma _{i1} +  \frac{ \langle  \gamma _i v _{ix} \rangle  }{ \vsh } \right)  \frac{ B _{x0} }{ B _0 }  .
\label{eq:a7}
\end{equation}
Because \( | v _{{i}x} | \) is smaller than \( \vsh \), the range of \( | \langle v _{{i}x} \rangle | \) is expected to be \( 1/2 \lto | \langle v _{{i}x} \rangle | / \vsh < 1 \), except maybe for extremely strong shock waves. In fact, if we assume a simple time dependence of \( v _{ix} \)
\begin{equation}
v _{ix} = - \vsh [ 1 - a _1 (t - t _1 )/(t _2 - t _1 ) ] \ \ \  \mbox{for} \ \ \  t _1 \le t \le t _2 ,
\label{eq:vix}
\end{equation}
with \( a _1 \) a constant (\( 0 \le a _1 \le 1 \)), then, after integrating over time, we have \( \langle v _{ix } \rangle = - \vsh  ( 1 - a _1  /2 ) \).
\par
\indent
The Lorentz factor for the fluid ion will be close to unity. Accordingly, for instance, if the ion fluid velocity \( v _{{i}x} \) is decreased to a value \(\sim - \vsh / 2 \), then the value of the content in the parentheses would be about 1/4.
\par
\indent
The electron speed will be \( \sim \vsh \) when the electrons encounter the shock and will be close to the speed of light when they are at the point \( x _{m}\). If we assume that the electron velocity is nearly parallel to the field line in the shock region, then (\( \vct{v} \cdot \drm \vct{b} / \drm t \)) can be neglected. Hence we have
\begin{equation}
\me ( - \gamma _{e2} c + \gamma _{e1} \vsh  B _{x0} / B _0  ) \simeq  - e E _{\parallel } ( t _2 - t _1  ) .
\label{eq:a8}
\end{equation}
The second term on the left-hand side, \( \gamma _{e1} \vsh B _{x0} / B _0  \), can be neglected compared with the first term \( \gamma _{e2} c \). The ions and electrons will have nearly the same time period (\( t _2 - t _1 \)) to keep charge neutrality. Substituting Eq. (\ref{eq:a7}) in (\ref{eq:a8}) yields 
\begin{equation}
\gamma _{e2} \simeq  \frac{ \mi \vsh }{ \me c } \left( \gamma _{i1} +  \frac{ \langle  \gamma _i v _{ix} \rangle }{ \vsh } \right)  \frac{ B _{x0} }{ B _0 }   .
\label{eq:a9}
\end{equation}
At \( x = x _{m} \), \( v_z \) can be estimated as \( v_z \simeq - c B _z / B \) (for reflected electrons the sign is reversed). For a large-amplitude shock, the ratio \( B _z / B \) is an order-unity quantity there. The quantity \( h \) defined by Eq. (\ref{eq:192}) is obtained as
\begin{equation}
h _2 \simeq \me c ^2 [ 1 +  \vsh B _{z0} / ( c B _{x0}  ) ]  .
\label{eq:a10}
\end{equation}
\par
\indent
If we average Eq. (\ref{eq:200}) over electrons in a small volume element, we have \begin{equation}
\langle h \gamma \rangle = e (F - F _0 ) + \me c ^2 \langle \gamma _{0} \rangle  .
\label{eq:a11}
\end{equation}
Here we used the relation \( \langle v _{z0} \rangle = 0 \). If electrons have the velocity \( v_z \simeq - c B _z / B \) at \( x = x _m \), then we find the maximum value of \( F \) from Eq. (\ref{eq:200}) as
\begin{equation}
e ( F _{m} - F _0 ) \simeq  h _2 \gamma _{e2} - \me c ^2 \langle \gamma _0 \rangle  .
\label{eq:a12}
\end{equation}
In many practical cases, we can take \( \langle \gamma _{0} \rangle \) to be order unity.
\par
\indent
Roughly speaking, the quantity \( e ( F _{m} - F _0 ) \) is of the order of \( \mi c \vsh \). It is, therefore, much greater than electron thermal energy \( \me \vte ^2 \) (\( \sim \mu B \)) and drift kinetic energy \( \me v _d  ^2 \) (\( \sim \me \vsh ^2 \)). 




%
%
\begin{figure}
\caption{Velocities and fields in the far upstream region in the wave frame. The velocity \( v _d  \) is \( E _{y0} \times B _0 \) drift, and \( - \vsh \) is equal to average guiding-center velocity in the \(x\) direction \( \langle v _{gx0} \rangle \). }
\label{fig:up}

\caption{Schematic diagram of magnetic field, electron velocity, \( \drm x \), and length \( \drm s \).}
\label{fig:ds}

\caption{Plot of function \( E _e ( v_{\parallel} ) = ( \me / 2 ) ( v_{\parallel} - v_{rv} ) ^2 + K \) at a fixed \(x \) position.}
\label{fig:ekin}

\caption{Schematic diagram of electron orbit in \( (x, \ y) \) plane.}
\label{fig:orb}

\caption{Schematic diagram of quantity [\( e ( F - F _0 ) + h _0 \gamma _0 \)]. In the top panel, this quantity is always positive. In the second one, it is negative in the dip. In the bottom one, it is always negative.}
\label{fig:fh}

\caption{Magnetic field profiles of an oblique shock at various times.}
\label{fig:mag}

\caption{Snapshots of field profiles. Electric and magnetic field profiles at \( \wpe t = 680 \) are plotted. They are normalized to \( B _0 \).}
\label{fig:eb}

\caption{Phase space plots of electrons.}
\label{fig:phase}

\caption{Profiles of \( E _{\parallel }, \ F \), and \( \varphi \) at two different 
times. Here, \( \tilde{ F } \) and  \( \tilde{ \varphi } \) denote \( e F / ( \me c ^2 ) \) and \( e \varphi / ( \me c ^2 ) \), respectively. }
\label{fig:efp}

\caption{Time variations of (\( x - v_{sh} t \) ), \( y \), \( z \), and \( \gamma \) of electrons. The lengths are normalized to the electron skin depth \( c / \wpe \).}
\label{fig:xyzg}

\caption{Maximum \( \gamma \) versus propagation angle \( \theta \). The propagation speeds for these shocks are \( \vsh \simeq 2.1 \valf \).}
\label{fig:angle}

\caption{ Maximum \( \gamma \) versus shock speed \( \vsh \). The propagation angle \( \theta \) is fixed to be \( \theta = 45 ^{\circ } \)}
\label{fig:vsh}
  
\end{figure}



%
%

%
%

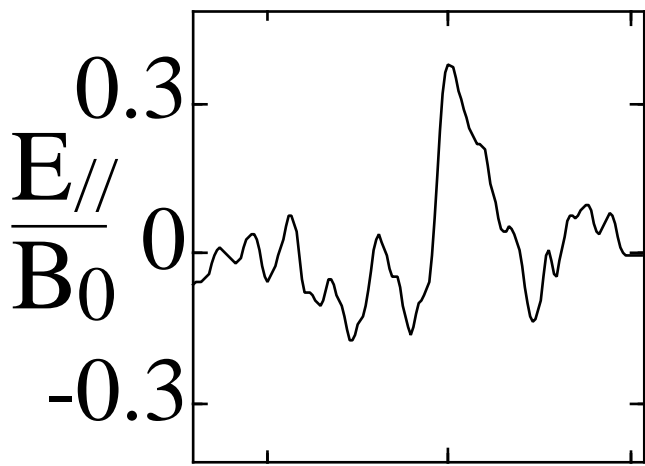
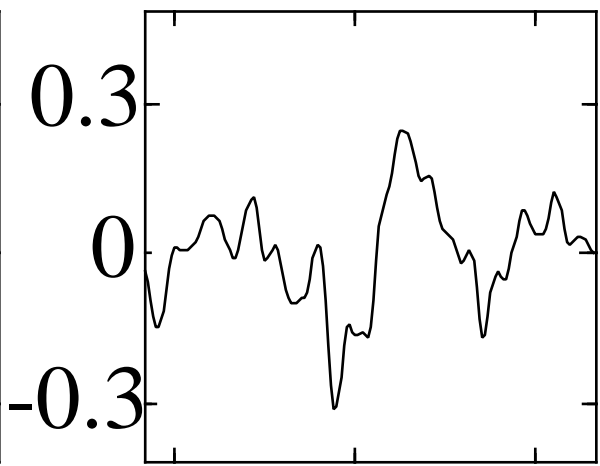
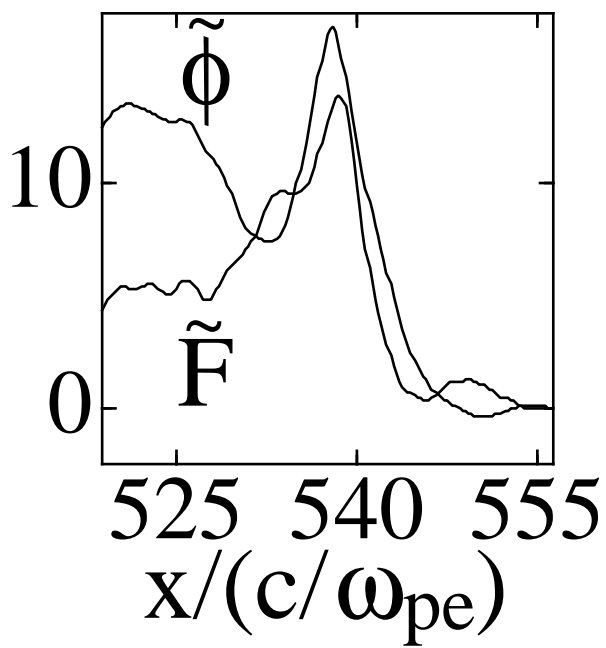
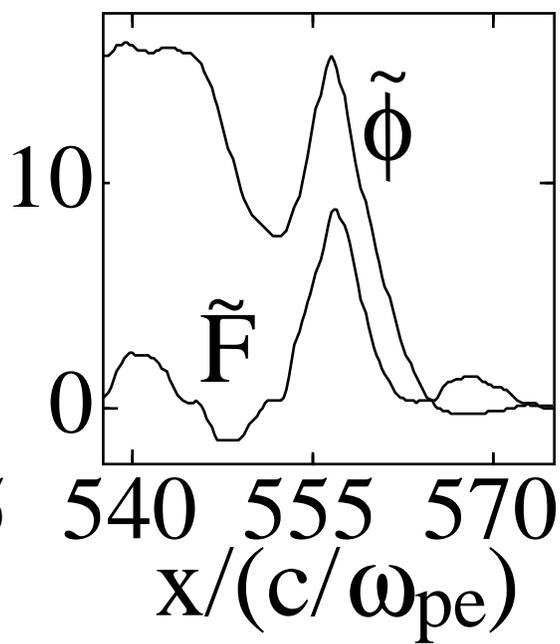
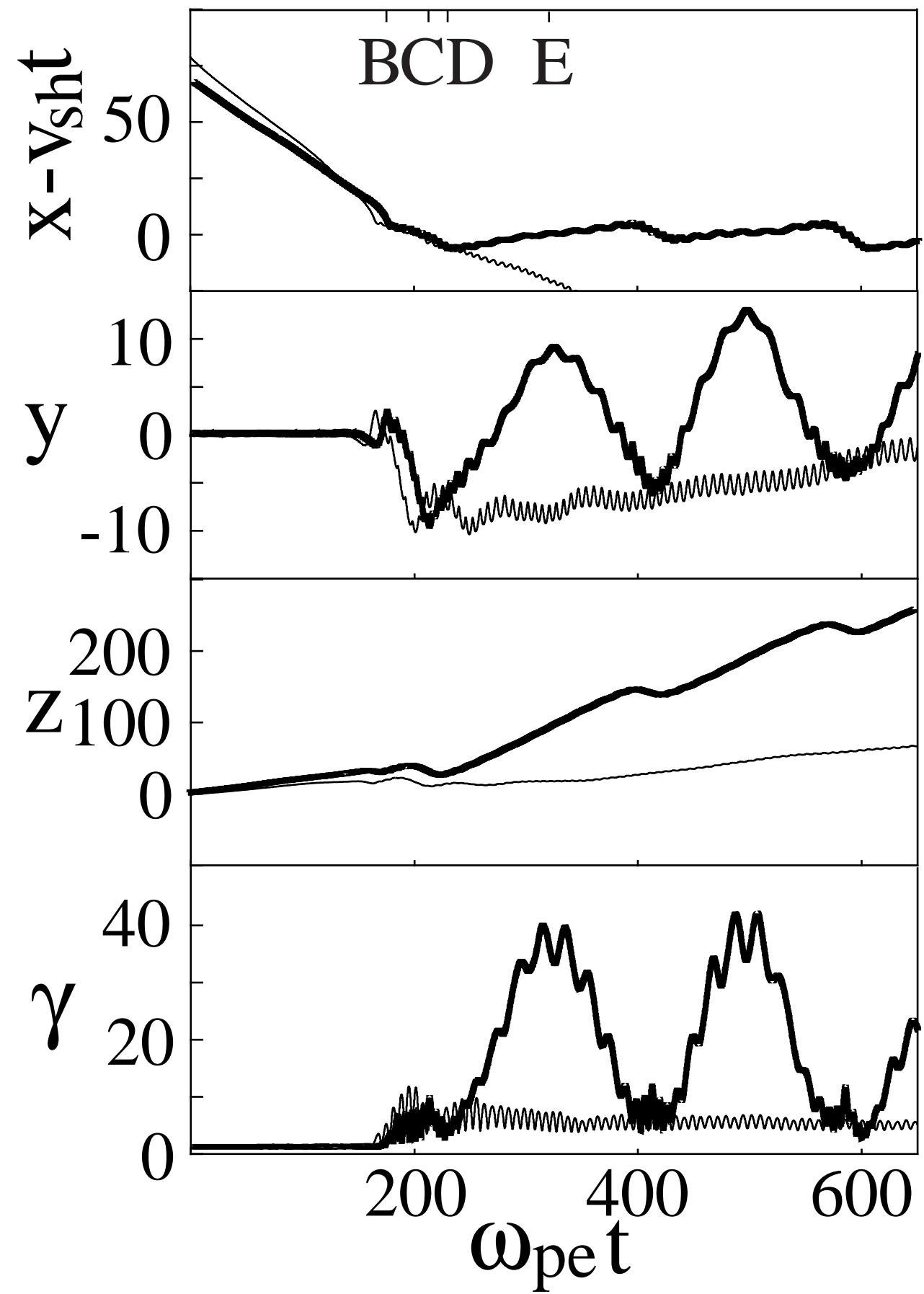
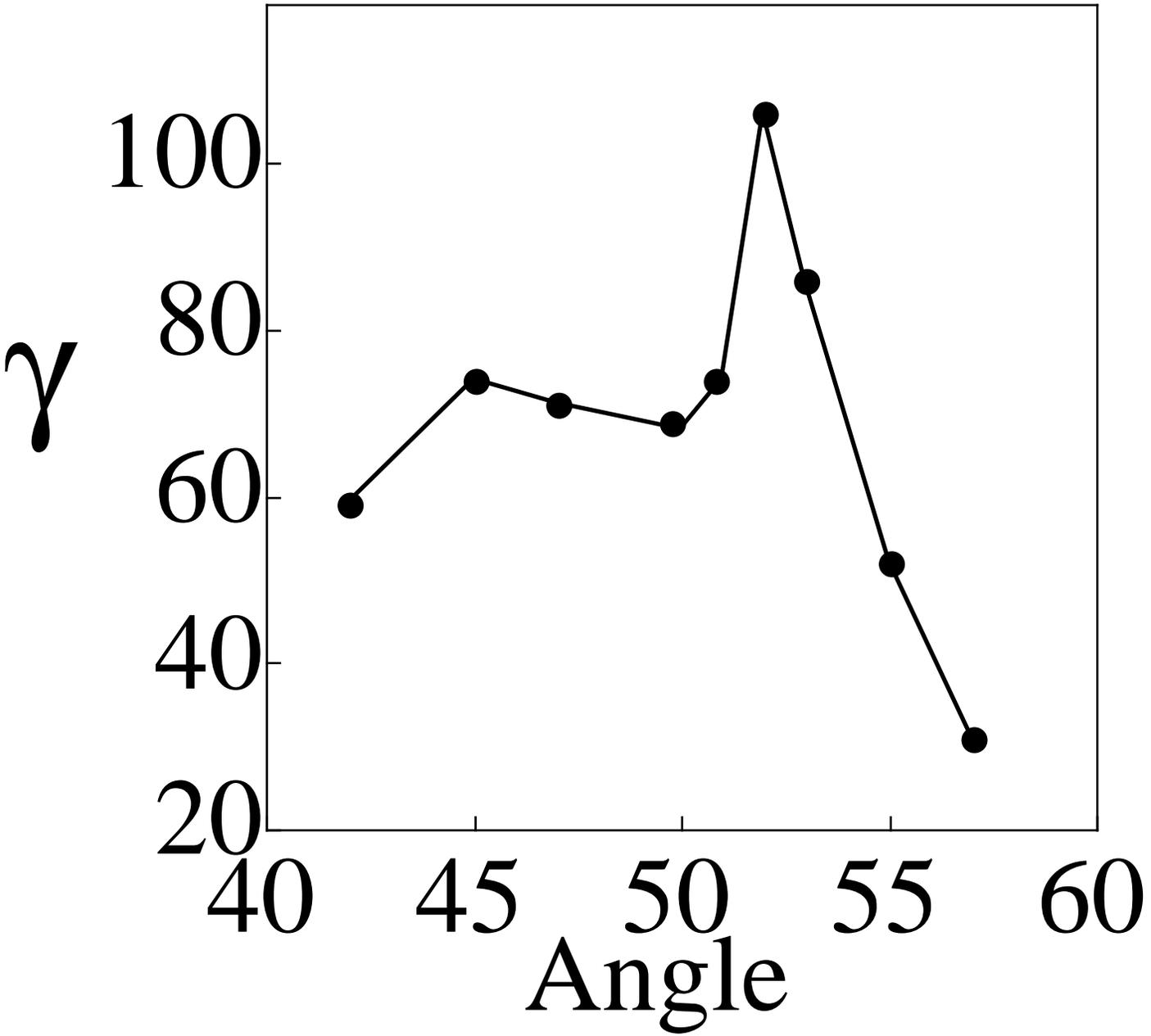
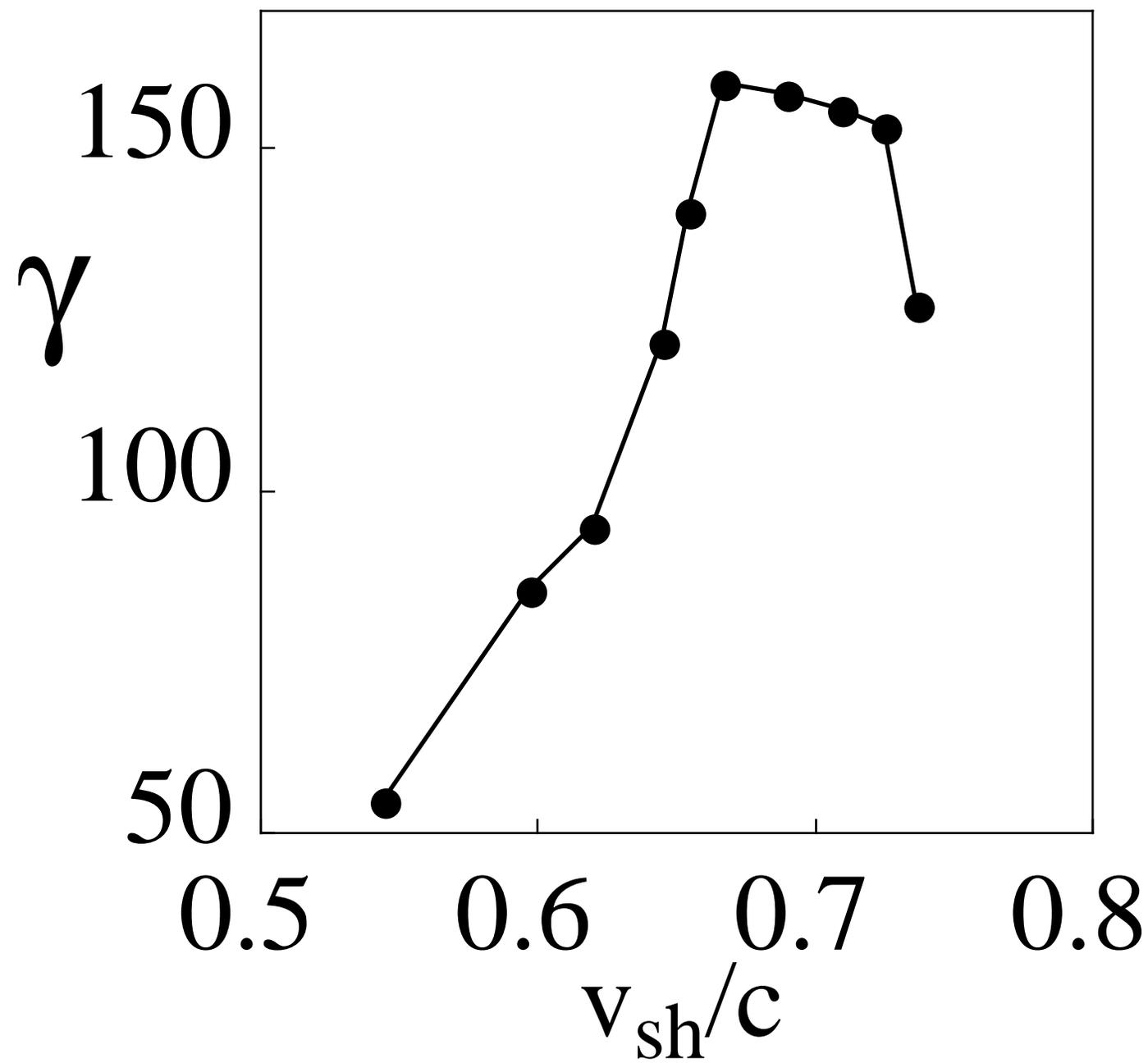
\end{document}